\documentclass[12pt]{article}

\usepackage{amsmath}
\usepackage{amssymb}
\usepackage{slashed}
\usepackage{amsfonts}
\usepackage{graphicx}
\usepackage[dvipsnames]{xcolor}
\usepackage[colorlinks=true,urlcolor=blue,linkcolor=blue,citecolor=blue]{hyperref}
\usepackage{cancel}
\usepackage{booktabs}
\usepackage{caption}
\usepackage{subcaption}
\usepackage{float}

\usepackage[top=1.135in,bottom=1.17in,left=1.16in,right=1.16in]{geometry}
\numberwithin{equation}{section}

\allowdisplaybreaks

\newcommand{\be}{\begin{equation}}
\newcommand{\ee}{\end{equation}}

\usepackage{graphicx} 
\usepackage{amssymb,amsthm,amsmath,amsfonts}
\usepackage[backend=biber, style=ieee]{biblatex}
\usepackage{makecell}
\usepackage{bm}
\usepackage{multicol}
\usepackage{appendix}

\usepackage[T1]{fontenc}

\begin{document}
\thispagestyle{empty}

\begin{center}

{\bf\Large \boldmath $\nu e\to\nu e$ scattering with massive Dirac or Majorana neutrinos and general interactions}

\vspace{50pt}

Juan Manuel Márquez, Pablo Roig and Mónica Salinas

\vspace{16pt}
{\it Departamento de F\'isica, Centro de Investigaci\'on y de Estudios Avanzados del Instituto Polit\'ecnico Nacional} \\
{\it Apartado Postal 14-740, 07000 Ciudad de México, M\'exico}
\vspace{16pt}

\vspace{50pt}

\vspace{16pt}
{\small
}
\vspace{16pt}

{\tt}

\today

\vspace{30pt}

\end{center}

\begin{abstract}
We calculate the neutrino-electron elastic scattering cross section, extending the results previously obtained in ref.~\cite{Rodejohann:2017vup}, in the presence of generic new interactions that take into account all the effects caused by finite neutrino masses. We address the potential significance of a heavy neutrino sector during precision measurements, particularly for tau neutrinos scattering with masses in the MeV range, for which the existing upper bounds on $|U_{\tau 4}|^2$ would result in conceivably measurable contributions. Finally, we comment on the possibility to distinguish between Dirac and Majorana neutrinos, including the analysis of the new emerging parameters and its application to illustrative model-dependent scenarios. 
\end{abstract}

\newpage


\section{Introduction}
Although the Standard Model of Elementary Particles (SM) has had a remarkable success at describing nature, remaining consistent with the experimental data \cite{pdg}, there are still some physical phenomena that motivate the search for new physics (NP) beyond the SM \cite{beyondSM}.

One of the most important open questions nowadays is whether neutrinos are Dirac or Majorana fermions, which is also crucial for understanding the origin of their masses and some interesting physical processes as allowed or forbidden. There are a number of proposals to probe the specific nature of the neutrino, being the most promising the neutrino-less double beta decay ($0\nu\beta\beta$) of nuclei \cite{DBeta1,DBeta2,DBeta3}\footnote{This could however be unobservable, even if neutrinos are Majorana particles, depending on their properties, like the neutrino mass ordering.}. There are alternative possibilities, such as coherent scattering of neutrinos on nucleus with bremsstrahlung radiation \cite{CoherentScat18}, lepton number violating processes \cite{LNV1,LNV2,LNV3,LNV4,LNV5,LNV6,LNV7}, lepton flavor changing decays \cite{Novales-Sanchez:2017crc}, electromagnetic properties of charged leptons \cite{Novales-Sanchez:2016sng}, and triple gauge boson couplings \cite{Martinez:2022epq,Novales-Sanchez:2023ztg},  most of them motivated by the propagation of massive heavy neutral leptons. Moreover, many searches for the neutrino's nature are also conducted within the framework of cosmology, including cosmic neutrino capture via tritium \cite{Long:2014zva}, the evaporation of primordial black holes \cite{Lunardini:2019zob}, the cosmological effective number of neutrino species \cite{Luo:2020fdt,Luo:2020sho} and more \cite{BahaBalantekin:2018ppj,Balantekin:2018ukw,Funcke:2019grs,SajjadAthar:2021prg}.

It is however extremely challenging to distinguish experimentally between Dirac and Majorana neutrinos, since any observable difference between both neutrinos nature is always suppressed by some power of $m_\nu/E$ in theories with $V-A$ interactions such as the SM \cite{Kayser:1982br}. Then, an alternative idea is to suppose that neutrinos have new interactions beyond the SM, where high precision measurements could help to discern the specific neutrino nature as well as the possible new physics involved.

In early works, the most general Lorentz-invariant form of neutrino-fermion interactions at operator dimension six was assumed, including scalar, pseudo-scalar, vector, axial-vector and tensor couplings \cite{Rosen:1982pj,Rodejohann:2017vup,Shrock:1981cq,Marquez:2022bpg,Doi:2005pm}. These, among many others, have pointed out that the energy spectrum of final charged leptons in their respective processes and the assumption of NP could be used to distinguish between Dirac and Majorana neutrinos and also to study the presence of a possible heavy neutrino sector. 

In this paper we generalize the work done in ref.~\cite{Rodejohann:2017vup}, computing the neutrino-electron elastic scattering cross section in the presence of general new interactions and including the effect of finite neutrino masses, where the presence of a heavy sector could be important while performing precision measurements. The general interaction case is discussed, as well as the specific results of selected model-dependent scenarios. 

This work is structured as follows: after a review of the effective neutrino-electron scattering Lagrangian in section \ref{sec:2}, we summarize the result of the cross section neglecting neutrino masses (section \ref{sec:2.1}) and present our new results considering finite neutrino masses (section \ref{sec:2.2}). Then, in section \ref{sec:3}, we discuss the possible effects of a heavy neutrino sector, specially in the case where the new sterile neutrinos have non-negligible mixing. After that, in section \ref{sec:4}, we analyse the possibility to distinguish between Dirac and Majorana neutrinos in this process, giving examples of its applicability to model-dependent scenarios. Finally, our conclusions are given in section \ref{sec:5}. Appendix \ref{appendix:A} includes an alternative notation for our general result that could be useful for future works.

\section{General neutrino-electron scattering interactions }
\label{sec:2}

Let us consider the most general Lorentz-invariant interaction between neutrinos and charged leptons pairs at dimension six~\cite{Rodejohann:2017vup}~\footnote{Ref.~\cite{Bischer:2019ttk} discusses its ultra-violet completion, from the effective field theory perspective.}:

\begin{align}
    \mathcal{L}\supset \frac{G_F}{\sqrt{2}} \sum_{a=S,P,V,A,T} \bar{\nu}\Gamma^a \nu \Bigl[ \bar{ l }\ \Gamma^a (C_a + \bar{D}_a i \gamma^5) l \Bigr],
    \label{lagrangiana}
\end{align}
where  $\Gamma^a$, with $a=S,P,V,A,T$; commonly known as scalar, pseudo-scalar, vector, axial-vector, and tensor interactions, respectively, are the several Dirac matrices that can be combined independently, 

\begin{align}
    \Gamma^a=\left\lbrace  I, i\gamma^5, \gamma^\mu, \gamma^\mu \gamma^5, \sigma^{\mu \nu} \equiv \frac{i}{2}[\gamma^\mu , \gamma^\nu] \right\rbrace.
\end{align}

Two options exist for contracting the terms in the tensor case, $g_{\mu \mu'} g_{\nu \nu'} \sigma^{\mu \nu} \sigma^{\mu' \nu'}$ and $\varepsilon_{\mu \nu \mu' \nu'}\sigma^{\mu \nu} \sigma^{\mu' \nu'}$. Following previous works, we only make use of the first one since, in accordance with reference \cite{Lindner:2016wff}, the latter can be turned into the former up to a redefinition of $C_T$ and $D_T$.

Hermiticity of eq.~(\ref{lagrangiana}) allows us to define
\begin{align}\label{Eq.Da&Dbara}
    &D_a \equiv \bar{D}_a, \quad  a=S,P,T,\\ \nonumber
    &D_a \equiv i\bar{D}_a, \quad  a=V,A,
\end{align}
thus, each coupling constant  $ C_a$ and $D_a$ is a real number.

Additionally, if we apply the Majorana condition $\nu_j=\nu_j^c= C\bar{\nu}_j^T$, where $C$ is the charge conjugation matrix, some coefficients must vanish ($C_V=D_V=C_T=D_T=0$). This is the reason why we will be able to distinguish between Dirac and Majorana cases in the general case.

As an example, the SM scenario is just a specific case of (\ref{lagrangiana}). Indeed the Lagrangian that describes the neutral current interaction (NC) in the SM is: \footnote{The charged current (CC) interaction may also contribute if the neutrino and the charged lepton have the same flavor. To include the charged current contribution one simply replaces $g_V^l \to g_V^l + 1$ and $g_A^l \to g_A^l + 1$ after a Fierz transformation.} 

\begin{align}
    \mathcal{L}_{\mathrm{NC}}= \frac{G_F}{\sqrt{2}}2\Bigl[\bar{\nu}\gamma^{\mu}(g_V^\nu-g_A^\nu \gamma^5)\nu\Bigr]\Bigl[\bar{l}\gamma^{\mu}(g_V^l-g_A^l \gamma^5)l\Bigr],
    \label{lagrangianaSM}
\end{align}
where
\begin{align}
        g_V^\nu=g_A^\nu=\frac{1}{2}, \quad g_V^l=-\frac{1}{2}+2s_\mathrm{w}^2, \quad g_A^l=-\frac{1}{2}.
        \label{SMconstants}
\end{align}

By comparing eqs.~(\ref{lagrangiana}) and (\ref{lagrangianaSM}), we get

\begin{align}
\footnotesize
\mathrm{\huge Dirac} \begin{cases}
C_{V}^{\mathrm{SM}}=2 g_V^\nu g_V^l, &   D_{V}^{\mathrm{SM}}=-2 g_V^\nu g_A^l,\\
\\
C_{A}^{\mathrm{SM}}=2 g_A^\nu g_A^l, &   D_{A}^{\mathrm{SM}}=-2 g_A^\nu g_V^l,\\
\\
C_{S}^{\mathrm{SM}}=0, &   D_{S}^{\mathrm{SM}}=0,\\
\\
C_{P}^{\mathrm{SM}}=0, &   D_{P}^{\mathrm{SM}}=0,\\
\\
C_{T}^{\mathrm{SM}}=0, &   D_{T}^{\mathrm{SM}}=0,
\end{cases}
\mathrm{\huge Majorana} \begin{cases}
C_{V}^{\mathrm{SM}}=0, &   D_{V}^{\mathrm{SM}}=0,\\
\\
C_{A}^{\mathrm{SM}}=4 g_A^\nu g_A^l, &   D_{A}^{\mathrm{SM}}=-4 g_A^\nu g_V^l,\\
\\
C_{S}^{\mathrm{SM}}=0, &   D_{S}^{\mathrm{SM}}=0,\\
\\
C_{P}^{\mathrm{SM}}=0, &   D_{P}^{\mathrm{SM}}=0,\\
\\
C_{T}^{\mathrm{SM}}=0, &   D_{T}^{\mathrm{SM}}=0.
\end{cases}
\label{diracSM}
\end{align}
We emphasize that for Majorana neutrinos $C_V=D_V=C_T=D_T=0$ and, for the specific SM case, the remaining ones double their values compared to the Dirac case.



\subsection{Cross section neglecting neutrino masses}
\label{sec:2.1}
As done in ref.~\cite{Rodejohann:2017vup}, the result for the cross section of elastic scattering of neutrinos (antineutrinos) with massive charged leptons at energies where the local interaction approximation holds, neglecting neutrino masses, is
\begin{equation}\label{eq.nu.massless}
\frac{d \sigma}{dT}(\nu + \textit{e})=\frac{G_F^2 M}{2 \pi }\left[ A + 2 B \left(1-\frac{T}{E_\nu}\right) +C\left(1-\frac{T}{E_\nu}\right)^2 + D\frac{MT}{4 E_\nu^2}   \right],
\end{equation}
\begin{equation}\label{eq.nubar.massless}
\frac{d \sigma}{dT}(\Bar{\nu} + \textit{e})=\frac{G_F^2 M}{2 \pi }\left[ C + 2 B \left(1-\frac{T}{E_\nu}\right) + A\left(1-\frac{T}{E_\nu}\right)^2 + D\frac{MT}{4 E_\nu^2}   \right],
\end{equation}
where $E_\nu$ is the incident neutrino energy, $T$ and $M$ are the recoil energy and mass of the charged lepton, respectively, and
\begin{align}
    A \equiv &\frac{1}{4}(C_A-D_A+C_V-D_V)^2+\frac{1}{8}(C_P^2 + C_S^2 + D_P^2 + D_S^2+ 8C_T^2 +8D_T^2)\nonumber\\ 
    &\quad +\frac{1}{2}(C_PC_T-C_SC_T+D_PD_T-D_SD_T),\\
    B\equiv &-\frac{1}{8}(C_P^2 + C_S^2 + D_P^2 + D_S^2- 8C_T^2 -8D_T^2),\\
    C \equiv &\frac{1}{4}(C_A+D_A-C_V-D_V)^2+\frac{1}{8}(C_P^2 + C_S^2 + D_P^2 + D_S^2+ 8C_T^2 +8D_T^2)\nonumber\\ 
    &-\frac{1}{2}(C_PC_T-C_SC_T+D_PD_T-D_SD_T),\\
    D \equiv& (C_A-D_V)^2-(C_V-D_A)^2-4(C_T^2+D_T^2)+C_S^2+D_P^2.
\end{align}
It is important to remember that all these calculations are derived under the assumption that the incoming neutrinos or anti-neutrinos are left-handed or right-handed, respectively. It is also specially interesting that both distributions are related by the exchange $A\leftrightarrow C$ and that none of the parameters mix the (axial)-vector type interactions with any other. Detailed explanations on this topic can be found in ref.~\cite{Rodejohann:2017vup}.

In this case, the SM values for the parameters $A,B,C,D
$, 
for NC contributions, are as follows
 \begin{align}\label{eq.parametersinSM}
     \Bigl(A,B,C,D\bigr)^{\mathrm{SM}}= \Bigl((1-2s_\mathrm{w}^2)^2,0,4s_\mathrm{w}^4,1-(1-4s_\mathrm{w}^2)^2\Bigr)\,,
 \end{align}
which lead to a cross section that has the same value for Dirac and Majorana neutrinos. Then, other interactions are needed in order to distinguish the Dirac and Majorana cases in this process. 

\subsection{Cross section considering neutrino masses}
\label{sec:2.2}
We now generalize the previous results considering finite neutrino masses and all their possible effects in the differential cross section, in order to further generalize the analysis.\\
If we consider finite neutrino masses, the current neutrino ($\nu_{L}$ for the weak isospin doublets electrically neutral component and $\nu_{R}'$ for singlets) is assumed to be the superposition of the mass-eigenstate neutrinos ($N_j$) with the mass $m_j$, that is,
\begin{equation}
    \nu_{\ell L}=\sum_j U_{\ell j}N_{jL},\quad \nu_{\ell R}'=\sum_j V_{\ell j}N_{jR},
\end{equation}
where $j=\{1,2,...,n\}$,  with $n$ the number of  mass-eigenstate neutrinos.

The $U$ matrix refers to the mixing of the active left-handed sector with massive neutrinos, while $V$ parameterizes the mixing of the sterile right-handed neutrinos. They are not directly related, except for the condition that unitarity or the specific model can implement.

For a complete discussion of this left- and right-handed mixing, we refer the reader to the appendix A of Ref.\cite{Doi:2005pm}, where it is explained in full detail, and matches our notation. For a complementary discussion, the lepton mixing formalism is also reviewed in the appendix A of Ref.\cite{Atre:2009rg}, with a slightly different notation but with great clarity.

 Thus, the explicit cross section, in the laboratory frame, of elastic neutrino (antineutrino) scattering on charged leptons is

\begin{align}
\frac{d \sigma}{dT}(\nu + \textit{e})=&\sum_{i,f}|U_{\ell i}|^2\frac{G_F^2 M}{2 \pi }\frac{E_\nu^2}{E_\nu^2-m_{\nu_i}^2}\left\lbrace A  + 2 B \left(1-\frac{T}{E_\nu}\right) +C\left(1-\frac{T}{E_\nu}\right)^2 \right.\nonumber  \\
&+ D \frac{MT}{4 E_\nu^2} +\frac{(m_{\nu_i}^2-m_{\nu_f}^2)}{2ME_\nu}\Bigl[(A+2B)+C\left(1-\frac{T}{E_\nu}\right)+ F \frac{m_{\nu_f}}{E_\nu}\Bigr]\nonumber\\
&-B \frac{m_{\nu_i}^2T}{ME_\nu^2}\left. + \frac{m_{\nu_f}}{E_\nu} \Bigl[ G + F \left(1-\frac{T}{E_\nu}\right)\Bigr]+ D \frac{m_{\nu_i}^2+m_{\nu_f}^2}{8E_\nu^2} \right\rbrace,
\label{completesec1}
\end{align}

\begin{align}
\frac{d \sigma}{dT}(\Bar{\nu} + \textit{e})=&\sum_{i,f}|U_{\ell i}|^2\frac{G_F^2 M}{2 \pi }\frac{E_\nu^2}{E_\nu^2-m_{\nu_i}^2}\left\lbrace C  + 2 B \left(1-\frac{T}{E_\nu}\right) +A\left(1-\frac{T}{E_\nu}\right)^2 \right.\nonumber  \\
&+ D \frac{MT}{4 E_\nu^2} +\frac{(m_{\nu_i}^2-m_{\nu_f}^2)}{2ME_\nu}\Bigl[(C+2B)+A\left(1-\frac{T}{E_\nu}\right)- G \frac{m_{\nu_f}}{E_\nu}\Bigr]\nonumber\\
&-B \frac{m_{\nu_i}^2T}{ME_\nu^2}\left. - \frac{m_{\nu_f}}{E_\nu} \Bigl[ F + G \left(1-\frac{T}{E_\nu}\right)\Bigr]+ D \frac{m_{\nu_i}^2+m_{\nu_f}^2}{8E_\nu^2} \right\rbrace,
\label{completesec2}
\end{align}
where $m_{\nu_{i}}$ ($m_{\nu_{f}}$) is the mass of the initial (final) neutrino and we only added two new parameters $F$ and $G$, that were not contributing in the massless neutrino case. These, together with the $A,\,B,\,C,\,D$ parameters already appearing in eqs.~(\ref{eq.nu.massless}) and (\ref{eq.nubar.massless}), are given, in the massive neutrino case, by
 \begin{align}
    A \equiv &|U_{\ell f}|^2\left[\frac{1}{4}(C_A-D_A+C_V-D_V)^2\right]+|V_{\ell f}|^2\left[\frac{1}{8}(C_P^2 + C_S^2 + D_P^2\right.\nonumber\\
    & + D_S^2+ 8C_T^2 +8D_T^2)\left.+\frac{1}{2}(C_PC_T-C_SC_T+D_PD_T-D_SD_T)\right],\\
    B\equiv &-|V_{\ell f}|^2\left[\frac{1}{8}(C_P^2 + C_S^2 + D_P^2 + D_S^2- 8C_T^2 -8D_T^2)\right],\\
    C \equiv &|U_{\ell f}|^2\left[\frac{1}{4}(C_A+D_A-C_V-D_V)^2\right]+|V_{\ell f}|^2\left[\frac{1}{8}(C_P^2 + C_S^2 + D_P^2\right.\nonumber\\
    & + D_S^2\left.+ 8C_T^2 +8D_T^2)-\frac{1}{2}(C_PC_T-C_SC_T+D_PD_T-D_SD_T)\right],\\
    D \equiv& |U_{\ell f}|^2\left[(C_A-D_V)^2-(C_V-D_A)^2\right]+|V_{\ell f}|^2\left[-4(C_T^2+D_T^2)\right.\nonumber\\
    &\left.+C_S^2+D_P^2\right],\\
    F\equiv & \operatorname{Re}\left[U_{\ell f}V^*_{\ell f}\right]\frac{1}{4}[(C_S +6C_T)(C_V-D_A)+(C_P -6C_T)(C_A-D_V)],\\
    G\equiv & \operatorname{Re}\left[U_{\ell f}V^*_{\ell f}\right]\frac{1}{4}[(C_S -6C_T)(C_V-D_A)-(C_P +6C_T)(C_A-D_V)].
\end{align}
The reason we have an overall factor of $|U_{\ell i}|^2$ in the differential cross section is because of the assumption that the incoming neutrinos or anti-neutrinos are left-handed or right-handed, respectively~\footnote{See, conversely, ref.\cite{Blaut:2018fis}.}; while the final neutrino could be produced in any chirality state, depending on the specific physics involved. We emphasize that the result is applicable to both relativistic neutrino scattering and the non-relativistic case, as we will discuss further later on.

Then, all these parameters can be extracted from scattering data and in principle could be used to explore the existence of a possible heavy neutrino sector as well as the specific nature of neutrinos, as we shall discuss in the next sections. It is also remarkable that this generalization of the cross section introduces only two more parameters, $F$ and $G$, which mix the vector and axial currents with the scalar, pseudoscalar and tensor ones; that is a new feature compared with the previous result \cite{Rodejohann:2017vup}. In the sections that follow, every element of these earlier findings will be clarified and examined. 

Finally, it is useful to remark some of the symmetry properties that the cross sections have. The cross sections for neutrinos and antineutrinos scattering are related by the exchange $A\leftrightarrow C$, $F\leftrightarrow G$, $m_{\nu_i}\leftrightarrow -m_{\nu_i}$ and $m_{\nu_f}\leftrightarrow -m_{\nu_f}$, which could be used while analyzing model-dependent scenarios. Also, in the light neutrino case, if we sum over all the possible mass-eigenstate neutrinos (using the unitarity condition) and neglect the explicit neutrino mass terms, we recover the results already obtained in ref.~\cite{Rodejohann:2017vup}, just as a trivial check of our expressions. 

\section{Signals from a heavy neutrino sector}
\label{sec:3}

One of the main goals of this computation is to include the effect of neutrino masses in order to study their possible contributions, specially on scenarios where the new heavy sterile neutrinos have non-negligible mixing with the light and active ones. 

The differential cross section obtained in eqs.(\ref{completesec1},\ref{completesec2}) is general under the discussed hypotheses, and in principle includes the contributions of all the neutrino mass eigenstates that can be produced kinematically. These have different neutrino mass dependence, such as: 
\begin{equation}
    \frac{m_{\nu}}{E_\nu},\quad \frac{m_{\nu}^2}{E^2_\nu},\quad \frac{m_{\nu}^2 T}{M E_\nu^2}.
    \label{eqn:massstructures}
\end{equation}

The above structures are precisely the ones that will suppress these contributions with respect to the terms without neutrino mass dependence. Nevertheless, for precision measurements, this suppression could be low enough to play a role in the energy spectrum, specifically under the assumption of a heavy neutrino sector, as we shall discuss. 

At this stage, we just want to motivate the presence of these new terms that appear in the differential decay rate due to accounting for finite neutrino masses. For this purpose we will begin with a rough, but still realistic, approximation. Since we are considering that the incident neutrinos (antineutrinos) are left (right)-handed, we will now assume that they are predominantly contributed by the light mass eigenstates, meaning that we will not have a suppression due to the global $|U_{\ell i}|^2$  mixing factor and $E_\nu^2-m_{\nu_i}^2\approx E_\nu^2$. Then, in order to have a sizable contribution of these new structures, we need to analyze the case of a heavy final state neutrino, where the suppression will come from its mixing factor $|U_{\ell f}|^2$\footnote{In the case of new physics, the mixing may be not suppressed, but the net suppression will be then encoded in the coupling. For this rough estimation we are not discussing all these possibilities.}.

In summary, we will now focus on a contribution with a light $m_{\nu_i}$ and a heavy $m_{\nu_f}$ with a suppressed mixing $|U_{\ell f}|^2$. In order to estimate the value of the structures shown in eq.~(\ref{eqn:massstructures}), we used the current experimental limits given in refs.~\cite{NA62:2017Electron, MicroBooNE:2023Muon,Barouki:2022bkt}. Then, we can estimate different cases, depending on the incident neutrino energy and the final neutrino mass. The findings of the suppression of the neutrino mass-dependent factors have been analyzed in table \ref{tab:suppression}, where we show the most relevant ones according to our estimation hypothesis and emphasize again that this is just a special case to motivate the presence of these terms.
\begin{table}[h!]
\begin{center}
\begin{tabular}{|c|c|c|c|c|}
\hline
 \makecell{Neutrino \\ flavor } & \makecell{$m_{\nu_f}$ \\ (MeV) }&  $|U_{\ell 4}|^2$ & \makecell{Linear term suppression \\ $|U_{\ell 4}|^2  m_{\nu_f} / E_\nu$}  &  \makecell{Quadratic term suppression \\ $|U_{\ell 4}|^2 m_{\nu_f}^2 / E_\nu^2$}\\
\hline
& & &\makecell{$E_\nu$ \\500 MeV}   \quad  \makecell{$E_\nu$  \\ 2500 MeV} & \makecell{$E_\nu$ \\500 MeV}   \quad  \makecell{$E_\nu$  \\ 2500 MeV} \\
\hline
\makecell{$l=e$ \\ \cite{NA62:2017Electron}} &\makecell{150-375\\ 375-440} & \makecell{$10^{-8}$ \\ $10^{-9}$} & \makecell{$10^{-9}$ \\ $10^{-10}$} \quad \makecell{ $10^{-10}$-$10^{-9}$ \\ $10^{-10}$ } & \makecell{ $10^{-10}$-$10^{-9}$ \\ $10^{-10}$} \quad \makecell{$10^{-10}$ \\ $10^{-11}$}\\
\hline
\makecell{$l=\mu$  \\ \cite{MicroBooNE:2023Muon}} &\makecell{10\\ 20 \\ 50 \\ 100 } & \makecell{$10^{-3}$ \\ $10^{-4}$ \\ $10^{-5}$ \\ $10^{-6}$} & \makecell{$10^{-5}$ \\ $10^{-6}$ \\ $10^{-6}$ \\ $10^{-7}$} \quad \makecell{$10^{-6}$ \\ $10^{-7}$ \\ $10^{-7}$ \\ $10^{-8}$} & \makecell{$10^{-7}$\\ $10^{-7}$ \\ $10^{-7}$ \\ $10^{-8}$} \quad \makecell{$10^{-8}$ \\ $10^{-9}$ \\ $10^{-9}$ \\ $10^{-9}$}  \\
\hline
\makecell{$l=\tau$  \\ \cite{Barouki:2022bkt}} & \makecell{ 100-200\\ 300-400 \\ 500-800 \\900-1100} & \makecell{$10^{-3}$ \\ $10^{-4}$ \\ $10^{-5}$ \\ $10^{-5}$} & \makecell{$10^{-4}$ \\ $10^{-4}$ \\ $10^{-5}$ \\ $10^{-5}$} \quad \makecell{$10^{-4}$ \\ $10^{-5}$ \\ $10^{-6}$ \\ $10^{-6}$} & \makecell{$10^{-4}$ \\ $10^{-4}$ \\ $10^{-5}$ \\ $10^{-5}$} \quad \makecell{$10^{-5}$ \\  $10^{-6}$-$10^{-5}$ \\ $10^{-6}$ \\ $10^{-6}$} \\
\hline
\end{tabular}
\end{center}
\caption{\label{tab:suppression} Estimated suppression associated with neutrino mass-dependent terms.}
\end{table}

As we can see, the suppression for electron neutrinos is really high because its mixing has tight constraints~\footnote{Even in this case, the contributions associated to the gauge boson longitudinal polarization can be neglected.}. The same happens with the muon neutrinos, which are not as constrained as the electron ones, but still too suppressed for the current experimental capabilities. Nevertheless, the limits on $|U_{\tau 4}|$ are the weakest, motivating the possibility that $|U_{\tau 4}|\gg|U_{e 5}|,|U_{\mu 4}|$ and thus getting suppression effects of the order $10^{-4}-10^{-5}$ which could produce measurable distortions in the energy spectrum, with the disadvantage that events with tau neutrinos are the rarest.

In this estimation, the linear neutrino mass terms with the presence of a heavy neutrino with a mass around $100-400$ MeV will be important for an incident neutrino energy of the order of $10^2-10^3$ MeV. It is also necessary to emphasize that these terms are always multiplied by the parameters $F$ and $G$, which are identically zero in the SM case, so we need the presence of new physics couplings in order to be sensitive to those contributions and then an extra suppression factor must be taken into account while analysing the possible distortions of the spectrum.

For the quadratic neutrino mass terms we also found interesting results. Indeed the suppression for the considered energy range could be of order $10^{-4}-10^{-5}$, which is a new feature compared to other analysis, see ergo \cite{Marquez:2022bpg}, where the quadratic terms were really suppressed. Also, these neutrino mass terms appear multiplying the $D$ parameter\footnote{They also appear in other parts of the spectrum, where the dependence is proportional to ($m_{\nu i}^2-m_{\nu f}^2$), so cancellations may occur while considering all the possible mass eigenstates, then we are not stating  further conclusions about this.}, which in eqs.~(\ref{eq.nu.massless}, \ref{eq.nubar.massless}) (neglecting neutrino masses) is suppressed by the electron mass with a factor $MT/E_\nu^2$. Then, depending on the recoil energy and the heavy neutrino sector, even the neutrino mass term could be of the same order as the one already obtained in ref.~\cite{Rodejohann:2017vup} for the $D$ parameter.

Here we recall that although our results apply to the non-relativistic case, the energy of the incident neutrino limits the possible mass of the heavy neutrino that could be accessed. Therefore, significant heavy neutrino effects may not be present in the low-energy regime. However, the non-relativistic case without neutrino mass effects is still interesting, as considered in Ref. \cite{Rodejohann:2017vup}.

Finally, more stringent conclusions could be obtained by restoring to model-dependent scenarios, where the full suppression could be estimated depending on the specific value of the new physics couplings. However, for our purpose, this rough estimation gives a clear idea of the possible signals of a heavy neutrino sector in this type of process. 

\section{Distinguishing Dirac from Majorana neutrinos}
\label{sec:4}

Our second goal is to use the information on the $A,\dots,G$ parameters, including the two new ones that appear while considering finite neutrino masses, and explore the corresponding parameter space allowed by Dirac or Majorana neutrinos. This way, it may be possible to distinguish between both neutrino natures depending on the values extracted from the experiment, as examined in ref.~\cite{Rodejohann:2017vup} for the case with negligible neutrino masses.

Applying the Majorana condition already discussed ($C_V=D_V=C_T=D_T=0$), we can obtain the explicit form of the parameters in the Dirac and Majorana cases, for the general interaction scenario, which are given in table \ref{tab:parameters}. This result will help us to quickly analyze many model-dependent scenarios, as we shall see.

 \begin{table}[h!]
\begin{center}
\begin{tabular}{|c|c|}
\hline
  Parameter & Dirac \\ [5pt]
\hline
$A$& \makecell{$|U_{\ell f}|^2\left[\frac{1}{4}(C_A-D_A+C_V-D_V)^2\right]+|V_{\ell f}|^2\left[\frac{1}{8}(C_P^2 + C_S^2 + D_P^2\right.$\\
     $+ D_S^2+ 8C_T^2 +8D_T^2)\left.+\frac{1}{2}(C_PC_T-C_SC_T+D_PD_T-D_SD_T)\right]$} \\ [10pt]
\hline
$B$&\makecell{ $-|V_{\ell f}|^2\left[\frac{1}{8}(C_P^2 + C_S^2 + D_P^2 + D_S^2- 8C_T^2 -8D_T^2)\right]$}\\ [10pt]
\hline
$C$& \makecell{$ |U_{\ell f}|^2\left[\frac{1}{4}(C_A+D_A-C_V-D_V)^2\right]+|V_{\ell f}|^2\left[\frac{1}{8}(C_P^2 + C_S^2 + D_P^2\right.$\\
$ + D_S^2\left.+ 8C_T^2 +8D_T^2)-\frac{1}{2}(C_PC_T-C_SC_T+D_PD_T-D_SD_T)\right]$}\\[10pt]
\hline
$D$& \makecell{$ |U_{\ell f}|^2\left[(C_A-D_V)^2-(C_V-D_A)^2\right]$\\
    $+|V_{\ell f}|^2\left[-4(C_T^2+D_T^2)+C_S^2+D_P^2\right]$}\\[10pt]
    \hline
$F$& \makecell{$ \operatorname{Re}\left[U_{\ell f}V^*_{\ell f}\right]\frac{1}{4}[(C_S +6C_T)(C_V-D_A)+(C_P -6C_T)(C_A-D_V)]$}\\[10pt]
\hline
$G$& \makecell{$\operatorname{Re}\left[U_{\ell f}V^*_{\ell f}\right]\frac{1}{4}[(C_S -6C_T)(C_V-D_A)-(C_P +6C_T)(C_A-D_V)]$}\\[10pt]
\hline
   & Majorana \\ [5pt]
\hline
$A$& \makecell{$|U_{\ell f}|^2\left[\frac{1}{4}(C_A-D_A)^2\right]+|V_{\ell f}|^2\left[\frac{1}{8}(C_P^2 + C_S^2 + D_P^2+ D_S^2)\right]$} \\ [10pt]
\hline
$B$&\makecell{ $-|V_{\ell f}|^2\left[\frac{1}{8}(C_P^2 + C_S^2 + D_P^2 + D_S^2\right]$}\\ [10pt]
\hline
$C$& \makecell{$ |U_{\ell f}|^2\left[\frac{1}{4}(C_A+D_A)^2\right]+|V_{\ell f}|^2\left[\frac{1}{8}(C_P^2 + C_S^2 + D_P^2+ D_S^2\right]$}\\
[10pt]
\hline
$D$& \makecell{$ |U_{\ell f}|^2\left[C_A^2-D_A^2\right]+|V_{\ell f}|^2\left[C_S^2+D_P^2\right]$}\\[10pt]
    \hline
$F$& \makecell{$ -\operatorname{Re}\left[U_{\ell f}V^*_{\ell f}\right]\frac{1}{4}[C_SD_A-C_PC_A]$}\\[10pt]
\hline
$G$& \makecell{$-\operatorname{Re}\left[U_{\ell f}V^*_{\ell f}\right]\frac{1}{4}[C_SD_A+C_PC_A]$}\\[10pt]
\hline
\end{tabular}
\end{center}
\caption{\label{tab:parameters} Parameters for Dirac and Majorana cases. The parameters for the latter case can be obtained from the former ones applying the properties explained in the paragraph below eq.~(\ref{Eq.Da&Dbara}).}
\end{table}
Let us start with the SM case (for NC contributions). Using eq.(\ref{SMconstants}) and eq.(\ref{diracSM}) in table \ref{tab:parameters}, we can compute the values for all the parameters in the SM scenario, that are shown in table \ref{tab:parametersSM}. 
 \begin{table}[h!]
\begin{center}
\begin{tabular}{|c|c|}
\hline
  Parameter & Dirac and Majorana \\ [5pt]
\hline
$A$& \makecell{$|U_{\ell f}|^2(1-2s_\mathrm{w}^2)^2$} \\ [10pt]
\hline
$B$&0\\ [10pt]
\hline
$C$& $|U_{\ell f}|^2 4s_\mathrm{w}^4$\\[10pt]
\hline
$D$& $|U_{\ell f}|^2\left(1-(1-4s_\mathrm{w}^2)^2\right)$\\[10pt]
    \hline
$F$& 0\\[10pt]
\hline
$G$&0\\[10pt]
\hline
\end{tabular}
\end{center}
\caption{\label{tab:parametersSM} Parameters for Dirac and Majorana cases in the SM, which are obtained from table \ref{tab:parameters} using eq.~(\ref{eq.parametersinSM}).}
\end{table}

Here it is straightforward to confirm that, if neutrino interactions are characterized by the SM, the cross section has the same value for Majorana and Dirac neutrinos, since all the parameters are equal in both cases. This is a result that still holds if we consider the effects of finite neutrino masses, which is an interesting feature, since in some other processes, specifically where both neutrinos appear in the final state, the difference between Dirac and Majorana cases is proportional to $m_\nu^2/E^2$ in the SM scenario due to the helicity flipping interference. 

Then, if only the SM interaction is involved in the $\nu e\to \nu e$ scattering, it is not possible in principle to distinguish the specific nature of neutrinos, even if their masses are taken into account. Nevertheless, as discussed in the last section, the presence of a heavy sector could be still possible to discriminate. Also, we can emphasize that in the SM case some parameters are identically zero; then, measuring a non-vanishing value for these parameters will imply the presence of new physics beyond the SM.

Now, we can give some other examples, including the presence of new physics couplings. The simplest case is to introduce just a new physics interaction beyond the SM. Considering the physical motivation in many model-dependent theories, we can analyse the case of SM interactions together with a non-zero scalar coupling $C_S$, that is shown in table \ref{tab:parametersSMS}.

 \begin{table}[h!]
\begin{center}
\begin{tabular}{|c|c|}
\hline
  Parameter & Dirac and Majorana \\ [5pt]
\hline
$A$& \makecell{$|U_{\ell f}|^2(1-2s_\mathrm{w}^2)^2+|V_{\ell f}|^2\frac{1}{8}C_S^2$} \\ [10pt]
\hline
$B$&$-|V_{\ell f}|^2\frac{1}{8}C_S^2$\\ [10pt]
\hline
$C$& $|U_{\ell f}|^2 4s_\mathrm{w}^4+|V_{\ell f}|^2\frac{1}{8}C_S^2$\\[10pt]
\hline
$D$& $|U_{\ell f}|^2\left(1-(1-4s_\mathrm{w}^2)^2\right)+|V_{\ell f}|^2 C_S^2$\\[10pt]
    \hline
$F$& $-\frac{1}{4}\operatorname{Re}\left[U_{\ell f}V^*_{\ell f}\right] C_S (1-4s_\mathrm{w}^2)$\\[10pt]
\hline
$G$& $-\frac{1}{4}\operatorname{Re}\left[U_{\ell f}V^*_{\ell f}\right] C_S (1-4s_\mathrm{w}^2)$ \\[10pt]
\hline
\end{tabular}
\end{center}
\caption{\label{tab:parametersSMS} Parameters for Dirac and Majorana cases in the SM$+C_S$ case.}
\end{table}
In this instance it is obvious that the presence of a new scalar coupling $C_S$ will imply a non-zero value of all the parameters compared to the SM case alone, which could be helpful to characterize the presence of this new physics. Nevertheless, the addition of just a $C_S$ coupling beyond the SM can not distinguish between Dirac and Majorana neutrinos, since the explicit dependence of the parameters in the couplings is the same for both cases.

Then, a model dependent scenario with a scalar sector involved would generate non-zero values for some parameters compared to the SM case and also would be able to access the new $F$ and $G$ parameters, which appear in the energy spectrum in the neutrino mass-dependent terms, helping us to study this kind of contributions. 

There could be many other scenarios, but for the Majorana condition in this process ($C_V=D_V=C_T=D_T=0$) it is clear that we need to introduce some of these vector or tensor couplings in order to get a testable difference between Dirac and Majorana neutrinos.

Then, as a final example, we analyse the case of a $C_T$ coupling beyond the SM case, which is of great interest since the $C_T$ coefficient must vanish for Majorana neutrinos, as we just discussed. The results are shown in table \ref{tab:parametersSMT}. 
 \begin{table}[h!]
\begin{center}
\begin{tabular}{|c|c|c|}
\hline
  Parameter & Dirac &Majorana \\ [5pt]
\hline
$A$& \makecell{$|U_{\ell f}|^2(1-2s_\mathrm{w}^2)^2+|V_{\ell f}|^2 C_T^2$} & $|U_{\ell f}|^2(1-2s_\mathrm{w}^2)^2$\\ [10pt]
\hline
$B$&$|V_{\ell f}|^2 C_T^2$&$0$\\ [10pt]
\hline
$C$& $|U_{\ell f}|^2 4s_\mathrm{w}^4+|V_{\ell f}|^2 C_T^2$&$|U_{\ell f}|^2 4s_\mathrm{w}^4$\\[10pt]
\hline
$D$& $|U_{\ell f}|^2\left(1-(1-4s_\mathrm{w}^2)^2\right)-4|V_{\ell f}|^2 C_T^2$&$|U_{\ell f}|^2\left(1-(1-4s_\mathrm{w}^2)^2\right)$\\[10pt]
    \hline
$F$& $6\operatorname{Re}\left[U_{\ell f}V^*_{\ell f}\right]C_T \ s_\mathrm{w}^2 $&$0$\\[10pt]
\hline
$G$& $3\operatorname{Re}\left[U_{\ell f}V^*_{\ell f}\right] C_T (1-2s_\mathrm{w}^2)$ &0\\[10pt]
\hline
\end{tabular}
\end{center}
\caption{\label{tab:parametersSMT} Parameters for Dirac and Majorana cases in the SM$+C_T$ case.}
\end{table}

As we can see, in this scenario, a new $C_T$ coupling would generate parameters that will differ for Dirac and Majorana neutrinos in all cases. Also, some of them must be identically zero for the Majorana case, so a measurement different from zero for the $B$, $F$ or $G$ parameters will imply that neutrinos are Dirac fermions if this model-dependent theory ($SM+C_T$) is the one describing nature. 

We can keep going on with all the possible new physics scenarios but as a final estimation we would like to follow the previous analysis \cite{Rodejohann:2017vup}, studying the parameter space for Dirac and Majorana neutrinos in the general case, now adding the information of the new $F$ and $G$ parameters.

As discussed in previous works (see e.g. \cite{Rodejohann:2017vup} and references therein), the parameters $A,B,C$ are the most accurate measurable quantities and then the most important ones for the current experiments, followed by the $D$ parameter, which is suppressed by the factor $M/E_\nu$. In this general analysis we have two more measurable parameters ($F, G$) that would be the most inaccurate among the six quantities because of the neutrino mass suppression already estimated. Nevertheless and for completeness of this work, we study the behavior of Dirac and Majorana neutrinos through the possible values of these new parameters in a similar way as done in ref.~\cite{Rodejohann:2017vup}. 

In \cite{Rodejohann:2017vup}, for the relativistic case, the normalized parameters triad ($A,B,C$) was studied randomly generating arbitrary values of ($C_a$, $D_a$) with $a=S, P, V, A$ and $T$, except that for Majorana neutrinos $C_V = D_V = C_T = D_T = 0$ were set, and then the corresponding ($A, B, C$) values were plotted and discussed, to see the accessible parameter space for Dirac and Majorana neutrinos. For the non-relativistic case, where the $D$ parameter is relevant, the four different parameters were studied similarly, in this case showing several 2-dimensional projections due to the extra parameter taken into account. 

In this general case we have now six parameters, so instead of showing several projections, we prefer to analyse the parameter space for some parameter triads in such a way that the subsequent discussions become clearer and generally encompass the distinction between Dirac and Majorana neutrinos. Thus, we study the parameter space ($X_1, X_2, X_3$), where $X_i$ could be any of the $A,\dots,G$ parameters, in the same way as \cite{Rodejohann:2017vup} randomly choosing values of ($C_a$, $D_a$) in $[-1,1]$. 

\begin{figure}[h!]
\begin{tabular}{cc}
 \includegraphics[scale=.65]{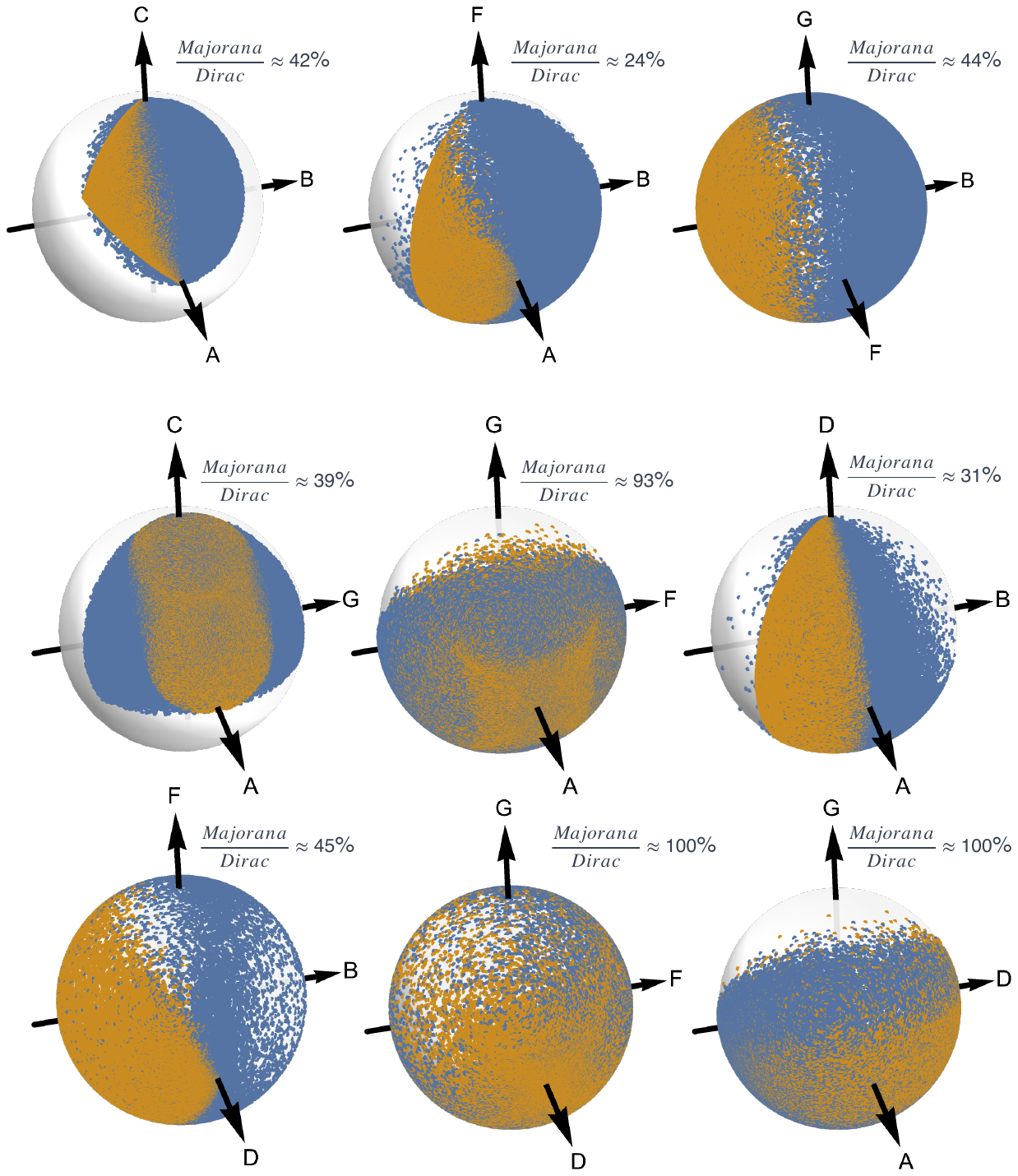} \\
 \end{tabular}
 \caption{Allowed values of different normalized triads of parameters ($X_1,X_2,X_3$) assuming Dirac (blue points) or Majorana (orange points) neutrinos. These plots were obtained with a random generation of \textbf{N} points inside the unit spheres as explained in the main text.}
 \label{fig:Pspace}
 \end{figure}

We also remember, as discussed in previous works, that for any allowed value of ($X_1, X_2, X_3$), ($r X_1, r X_2, r X_3$) is also allowed for any positive $r$, since it just corresponds to a rescaling of ($C_a$, $D_a$) by a factor of $\sqrt{r}$. Then, for simplicity, we normalize ($X_1, X_2, X_3$) to show the allowed region as usually done. Thus, figure \ref{fig:Pspace} shows the values of $10^5$ samples generated by randomly choosing values of ($C_a$, $D_a$) in $[-1,1]$ for different parameter triads ($X_1, X_2, X_3$). 
 
The difference between the Dirac (blue points) and Majorana (orange points) regions is evident in many of them, while other triads are not capable of distinguishing between Dirac and Majorana neutrinos on their own, since their boundaries are the same for both cases. In a similar way as the criterion proposed by Rosen, if ($X_1, X_2, X_3$) were measured in the region outside the orange space, then neutrinos would be Dirac particles, but if ($X_1, X_2, X_3$) were measured where blue and orange points overlap, then both Dirac and Majorana nature would be possible. 

Indeed, we estimate the common region of Dirac and Majorana points for each plot, giving an approximate percentage of the overlapping areas (with estimated relative uncertainty of $\sim2.5\%$). We accomplished this by assigning a surface to each case (Dirac and Majorana) and computing their area to determine the extent of overlap. For example, this implies that if the overlapping percentage is 100$\%$, we cannot distinguish between Dirac and Majorana nature, as both cases share the same parameter space boundaries. This is just a first estimation where the corresponding reported percentages may exhibit slight variations depending on the numerical analysis but it is still a helpful feature for further discussions in distinguishing the corresponding neutrino nature within a specific set of parameters.

The first triad ($A,B,C$) reproduces the result obtained in \cite{Rodejohann:2017vup}, where the complete discussions can be found, being the most promising way to differentiate the neutrino nature in this process. Additionally, we show eight more possible triads, which help us to illustrate some important aspects.

First, we have two possibilities: Either the Dirac and Majorana boundaries for the corresponding triads are different and then we clearly have a distinction between blue and orange points; which holds for the $(A,B,C)$, $(A,B,F)$, $(F,B,G)$, $(A,G,C)$, $(A,B,D)$ and $(D,B,F)$ triads. Or, conversely, that the Dirac and Majorana boundaries are the same, as the $(A,F,G)$, $(D,F,G)$ and $(A,D,G)$ cases, in which we are unable to distinguish the specific neutrino nature (assuming that only the corresponding triad could be measured), as the overlapping percentages suggest. 

The corresponding boundaries depend explicitly on the parameters coupling dependence. As an easy example \cite{Rodejohann:2017vup} 
we have the $B$ parameter, which has the following dependence for the Dirac and Majorana cases:
\begin{equation}
\begin{split}
     &B=-\frac{1}{8}(C_P^2 + C_S^2 + D_P^2 + D_S^2)+ C_T^2 +D_T^2 \quad \text{(Dirac)},\\
     & B=-\frac{1}{8}(C_P^2 + C_S^2 + D_P^2 + D_S^2) \quad\quad\quad\quad\quad\quad \ \text{(Majorana),}
\end{split}
\end{equation}
whence is evident that $B\leq 0$ for the Majorana case, while for the Dirac case it can have either sign. This is precisely why we have a clear distinction between blue and orange points in every triad in which $B$ appears. This analysis can be applied to any other parameter, the conclusion being less direct because of the corresponding explicit dependence.

We see now that the addition of these two new parameters $F$ and $G$ could give complementary information about the neutrino nature as well as of the possible heavy sector. Even if in some cases, such as the $(A,F,G)$ or $(D,F,G)$ triads, the accessible parameters region is not helpful to distinguish the corresponding neutrino nature; there are some others, such as the triad ($A,B,F$)
, in which the difference could be evident, giving more information than the case of negligible neutrino masses. 

For a final remark, as mentioned and justified in ref.~\cite{Rodejohann:2017vup}, the triad ($A, B, C$) remains the most important, because these parameters do not have an extra suppression in the effective cross section and their extraction from experimental data is much cleaner and more precise, giving the strongest constraints. Nevertheless, the analysis of $F$ and $G$ was also presented here for completeness and to illustrate the neutrino nature effect in each parameter, as discussed in this whole section.

\section{Conclusions}
\label{sec:5}

We still not know whether the observed neutrinos are Dirac or Majorana particles, or if the lepton sector includes additional fermion singlets (sterile neutrinos). The specific nature of neutrinos, as well as the existence of sterile neutrinos and new physics couplings, would affect the $\nu e\to\nu e$ scattering in a non-trivial way.

We have calculated the neutrino-electron elastic scattering cross section in the presence of general new interactions including all the effects due to finite neutrino masses, generalizing in this way the results obtained in ref.~\cite{Rodejohann:2017vup}. 

We have introduced two new parameters that arise due to considering finite neutrino masses and studied the effects of a possible heavy neutrino sector with a non-negligible mixing, as well as the impact of the specific neutrino nature on the differential cross section. 

We have found for our first estimation, using current experimental constraints on an invisible heavy neutrino, that there is some allowed heavy masses range, together with a non-negligible heavy-light mixing, that could lead to measurable contributions. Specifically, for the case of a tau neutrino dispersion with a mass around $100-400$ MeV and an incident neutrino energy on the ballpark of $10^{2}-10^{3}$ MeV, the linear term suppression could be of order $10^{-4}-10^{-5}$, which is a shared feature with the quadratic neutrino-mass term, unlike the results obtained on analogous processes where its suppression was very large. We motivate in this way the search for deviations from the SM in these processes, that might be observed if higher precision is achieved by future neutrino scattering experiments.

We have also discussed the possible ways to distinguish between Dirac and Majorana neutrinos in this particular process, giving some model-dependent examples where the new physics involved would generate a different parameter behaviour depending on the neutrino nature. We ended with a general numerical analysis adding the information of the new parameters introduced ($F,G$), motivated by the analogous study in the massless neutrino case, performed in ref.~\cite{Rodejohann:2017vup}.

Altogether, this reaffirms the importance of the search for new neutrino interactions, as well as for the presence of a heavy sector, which is surely of great interest, abounding in this  exciting field as a hot topic.

\section*{Acknowledgments}
J. M. is indebted to Conahcyt funding his Ph. D. P. R. thanks Conahcyt funds.
M. S. acknowledges funding by Conahcyt, through the program "Estancias Posdoctorales por México". We are grateful to Gabriel L\'opez Castro for useful comments and discussions on this work.

\appendix
\section{Alternative notation}
\label{appendix:A}
It is possible to write our general results using a different notation that may be useful: 

\begin{align}
\frac{d \sigma}{dT}(\nu + \textit{e})=&\sum_{i,f}\frac{G_f^2 M}{2 \pi }\frac{E_\nu^2}{E_\nu^2-m_{\nu_i}^2}\left\lbrace A  + 2 B \left(1-\frac{T}{E_\nu}\right) +C\left(1-\frac{T}{E_\nu}\right)^2 \right.\nonumber  \\
&+ D \frac{MT}{4 E_\nu^2} +\frac{(m_{\nu_i}^2-m_{\nu_f}^2)}{2ME_\nu}\Bigl[(A+2B)+C\left(1-\frac{T}{E_\nu}\right)+ F \frac{m_{\nu_f}}{E_\nu}\Bigr]\nonumber\\
&-B \frac{m_{\nu_i}^2T}{ME_\nu^2}\left. + \frac{m_{\nu_f}}{E_\nu} \Bigl[ G + F \left(1-\frac{T}{E_\nu}\right)\Bigr]+ D \frac{m_{\nu_i}^2+m_{\nu_f}^2}{8E_\nu^2} \right\rbrace,
\end{align}

\begin{align}
\frac{d \sigma}{dT}(\Bar{\nu} + \textit{e})=&\sum_{i,f}\frac{G_f^2 M}{2 \pi }\frac{E_\nu^2}{E_\nu^2-m_{\nu_i}^2}\left\lbrace C  + 2 B \left(1-\frac{T}{E_\nu}\right) +A\left(1-\frac{T}{E_\nu}\right)^2 \right.\nonumber  \\
&+ D \frac{MT}{4 E_\nu^2} +\frac{(m_{\nu_i}^2-m_{\nu_f}^2)}{2ME_\nu}\Bigl[(C+2B)+A\left(1-\frac{T}{E_\nu}\right)- G \frac{m_{\nu_f}}{E_\nu}\Bigr]\nonumber\\
&-B \frac{m_{\nu_i}^2T}{ME_\nu^2}\left. - \frac{m_{\nu_f}}{E_\nu} \Bigl[ F + G \left(1-\frac{T}{E_\nu}\right)\Bigr]+ D \frac{m_{\nu_i}^2+m_{\nu_f}^2}{8E_\nu^2} \right\rbrace,
\end{align}

 where
 \begin{align}
    A \equiv &\frac{1}{4}|C_A-D_A+C_V-D_V|^2+\frac{1}{8}(|C_P|^2 + |C_S|^2 + |D_P|^2 + |D_S|^2 \nonumber\\
    &+8|C_T|^2 +8|D_T|^2)+\frac{1}{2}(C_PC^*_T-C_SC^*_T+D_PD^*_T-D_SD^*_T),\\
    B\equiv &-\frac{1}{8}(|C_P|^2 + |C_S|^2 + |D_P|^2 + |D_S|^2- 8|C_T|^2 -8|D_T|^2),\\
    C \equiv &\frac{1}{4}|C_A+D_A-C_V-D_V|^2+\frac{1}{8}(|C_P|^2 + |C_S|^2 + |D_P|^2 + |D_S|^2\nonumber\\
    &+ 8|C_T|^2 +8|D_T|^2)-\frac{1}{2}(C_PC^*_T-C_SC^*_T+D_PD^*_T-D_SD^*_T),\\
    D \equiv& |C_A-D_V|^2-|C_V-D_A|^2-4(|C_T|^2+|D_T|^2)+|C_S|^2+|D_P|^2,\\
    F\equiv & \frac{1}{4}\operatorname{Re}[(C_S +6C_T)(C_V-D_A)^*+(C_P -6C_T)(C_A-D_V)^*],\\
    G\equiv & \frac{1}{4}\operatorname{Re}[(C_S -6C_T)(C_V-D_A)^*-(C_P +6C_T)(C_A-D_V)^*],
\end{align}
with
\begin{equation}
    C_a\equiv \left\lbrace\begin{array}{c} U_{\ell i}U^*_{\ell f} \bm{C_a} \quad \text{if    a=V,A,} \\ U_{\ell i}V^*_{\ell f}\bm{C_a} \quad \text{if    a=S,P,T,}\end{array}\right.
\end{equation}
\begin{equation}
    D_a\equiv \left\lbrace\begin{array}{c} U_{\ell i}U^*_{\ell f}\bm{D_a} \quad \text{if    a=V,A,} \\ U_{\ell i}V^*_{\ell f}\bm{D_a} \quad \text{if    a=S,P,T,}\end{array}\right.
\end{equation}
where the (real) constants in bold ($\bm{C_a}$ and $\bm{D_a}$) are those defined at the beginning of this work, eqs.~(\ref{lagrangiana}) and (\ref{Eq.Da&Dbara}). \\
This notation may look better, the price to pay is that now the new constants, $C_a$ and $D_a$, are complex. We just address this possibility here, that may be used in future works. 
\printbibliography

@article{pdg,
    author = "Workman, R. L. and Others",
    collaboration = "Particle Data Group",
    title = "{Review of Particle Physics}",
    doi = "10.1093/ptep/ptac097",
    journal = "PTEP",
    volume = "2022",
    pages = "083C01",
    year = "2022"
}

@misc{beyondSM,
  title = {Phenomenology of Physics beyond the Standard Model},
  author = {Kai Schmidt-Hoberg},
  note  =  {\url{https://www.desy.de/~kschmidt/BSM_lecturenotes.pdf}},
 }

@article{DBeta1,
    author = "Rodejohann, Werner",
    title = "{Neutrino-less Double Beta Decay and Particle Physics}",
    eprint = "1106.1334",
    archivePrefix = "arXiv",
    primaryClass = "hep-ph",
    doi = "10.1142/S0218301311020186",
    journal = "Int. J. Mod. Phys. E",
    volume = "20",
    pages = "1833--1930",
    year = "2011"
}

@article{DBeta2,
    author = "Vergados, J. D. and Ejiri, H. and Simkovic, F.",
    title = "{Theory of Neutrinoless Double Beta Decay}",
    eprint = "1205.0649",
    archivePrefix = "arXiv",
    primaryClass = "hep-ph",
    doi = "10.1088/0034-4885/75/10/106301",
    journal = "Rept. Prog. Phys.",
    volume = "75",
    pages = "106301",
    year = "2012"
}

@article{DBeta3,
    author = "Dolinski, Michelle J. and Poon, Alan W. P. and Rodejohann, Werner",
    title = "{Neutrinoless Double-Beta Decay: Status and Prospects}",
    eprint = "1902.04097",
    archivePrefix = "arXiv",
    primaryClass = "nucl-ex",
    doi = "10.1146/annurev-nucl-101918-023407",
    journal = "Ann. Rev. Nucl. Part. Sci.",
    volume = "69",
    pages = "219--251",
    year = "2019"
}

@article{CoherentScat18,
    author = "Millar, Alexander and Raffelt, Georg and Stodolsky, Leo and Vitagliano, Edoardo",
    title = "{Neutrino mass from bremsstrahlung endpoint in coherent scattering on nuclei}",
    eprint = "1810.06584",
    archivePrefix = "arXiv",
    primaryClass = "hep-ph",
    reportNumber = "MPP-2018-245, NORDITA-2018-095",
    doi = "10.1103/PhysRevD.98.123006",
    journal = "Phys. Rev. D",
    volume = "98",
    number = "12",
    pages = "123006",
    year = "2018"
}

@article{LNV1,
    author = "Hern\'andez-Tom\'e, G. and Illana, J. I. and Masip, M. and L\'opez Castro, G. and Roig, P.",
    title = "{Effects of heavy Majorana neutrinos on lepton flavor violating processes}",
    eprint = "1912.13327",
    archivePrefix = "arXiv",
    primaryClass = "hep-ph",
    doi = "10.1103/PhysRevD.101.075020",
    journal = "Phys. Rev. D",
    volume = "101",
    number = "7",
    pages = "075020",
    year = "2020"
}

@article{LNV2,
    author = "Hern\'andez-Tom\'e, G. and Castro, G. L\'opez and Portillo-S\'anchez, D.",
    title = "{\ensuremath{\Delta}L=2 hyperon decays induced by Majorana neutrinos and doubly charged scalars}",
    eprint = "2112.02227",
    archivePrefix = "arXiv",
    primaryClass = "hep-ph",
    doi = "10.1103/PhysRevD.105.113001",
    journal = "Phys. Rev. D",
    volume = "105",
    number = "11",
    pages = "113001",
    year = "2022"
}

@article{LNV3,
    author = "Hern\'andez-Tom\'e, Gerardo and Portillo-S\'anchez, Diego and Toledo, Genaro",
    title = "{Resonant Majorana neutrino effects in \ensuremath{\Delta}L=2 four-body hyperon decays}",
    eprint = "2212.03994",
    archivePrefix = "arXiv",
    primaryClass = "hep-ph",
    doi = "10.1103/PhysRevD.107.055042",
    journal = "Phys. Rev. D",
    volume = "107",
    number = "5",
    pages = "055042",
    year = "2023"
}

@article{LNV4,
    author = "Ilakovac, A.",
    title = "{Probing lepton number / flavor violation in semileptonic $\tau$ decays into two mesons}",
    eprint = "hep-ph/9608218",
    archivePrefix = "arXiv",
    reportNumber = "ZTF-95-06",
    doi = "10.1103/PhysRevD.54.5653",
    journal = "Phys. Rev. D",
    volume = "54",
    pages = "5653--5673",
    year = "1996"
}

@article{LNV5,
    author = "Lopez Castro, Gabriel and Quintero, Nestor",
    title = "{Lepton number violating four-body tau lepton decays}",
    eprint = "1203.0537",
    archivePrefix = "arXiv",
    primaryClass = "hep-ph",
    doi = "10.1103/PhysRevD.85.076006",
    journal = "Phys. Rev. D",
    volume = "85",
    pages = "076006",
    year = "2012",
    note = "[Erratum: Phys.Rev.D 86, 079904 (2012)]"
}

@article{LNV6,
    author = "Kim, C. S. and L\'opez Castro, G. and Sahoo, Dibyakrupa",
    title = "{Discovering intermediate mass sterile neutrinos through $\tau^- \to \pi^- \mu^- e^+ \nu$ (or $\bar{\nu}$) decay}",
    eprint = "1708.00802",
    archivePrefix = "arXiv",
    primaryClass = "hep-ph",
    doi = "10.1103/PhysRevD.96.075016",
    journal = "Phys. Rev. D",
    volume = "96",
    number = "7",
    pages = "075016",
    year = "2017"
}

@article{LNV7,
    author = "Quintero, N. and Lopez Castro, G. and Delepine, D.",
    title = "{Lepton number violation in top quark and neutral B meson decays}",
    eprint = "1108.6009",
    archivePrefix = "arXiv",
    primaryClass = "hep-ph",
    doi = "10.1103/PhysRevD.84.096011",
    journal = "Phys. Rev. D",
    volume = "84",
    pages = "096011",
    year = "2011",
    note = "[Erratum: Phys.Rev.D 86, 079905 (2012)]"
}

@article{Novales-Sanchez:2017crc,
    author = "Novales-S\'anchez, H. and Salinas, M. and Toscano, J. J.",
    title = "{About heavy neutrinos: Lepton-flavor violation in decays of charged leptons}",
    eprint = "1710.08474",
    archivePrefix = "arXiv",
    primaryClass = "hep-ph",
    doi = "10.1088/1361-6471/aad53c",
    journal = "J. Phys. G",
    volume = "45",
    number = "9",
    pages = "095004",
    year = "2018"
}

@article{Novales-Sanchez:2016sng,
    author = "Novales-S\'anchez, H. and Salinas, M. and Toscano, J. J. and V\'azquez-Hern\'andez, O.",
    title = "{Electric dipole moments of charged leptons at one loop in the presence of massive neutrinos}",
    eprint = "1610.06649",
    archivePrefix = "arXiv",
    primaryClass = "hep-ph",
    doi = "10.1103/PhysRevD.95.055016",
    journal = "Phys. Rev. D",
    volume = "95",
    number = "5",
    pages = "055016",
    year = "2017"
}

@article{Martinez:2022epq,
    author = "Mart\'\i{}nez, Eduardo and Monta\~no-Dom\'\i{}nguez, Javier and Novales-S\'anchez, H\'ector and Salinas, M\'onica",
    title = "{New physics in WW\ensuremath{\gamma} at one loop via Majorana neutrinos}",
    eprint = "2211.04629",
    archivePrefix = "arXiv",
    primaryClass = "hep-ph",
    doi = "10.1103/PhysRevD.107.035025",
    journal = "Phys. Rev. D",
    volume = "107",
    number = "3",
    pages = "035025",
    year = "2023"
}

@article{Novales-Sanchez:2023ztg,
    author = "Novales-S\'anchez, H\'ector and Salinas, M\'onica",
    title = "{Majorana neutrinos in the triple gauge boson coupling ZZZ*}",
    eprint = "2309.02400",
    archivePrefix = "arXiv",
    primaryClass = "hep-ph",
    doi = "10.1103/PhysRevD.108.075032",
    journal = "Phys. Rev. D",
    volume = "108",
    number = "7",
    pages = "075032",
    year = "2023"
}

@article{Rosen:1982pj,
    author = "Rosen, Simon Peter",
    title = "{Analog of the Michel Parameter for Neutrino - Electron Scattering: A Test for Majorana Neutrinos}",
    reportNumber = "PURD-TH-81-8",
    doi = "10.1103/PhysRevLett.48.842",
    journal = "Phys. Rev. Lett.",
    volume = "48",
    pages = "842",
    year = "1982"
}

@article{Rodejohann:2017vup,
    author = "Rodejohann, Werner and Xu, Xun-Jie and Yaguna, Carlos E.",
    title = "{Distinguishing between Dirac and Majorana neutrinos in the presence of general interactions}",
    eprint = "1702.05721",
    archivePrefix = "arXiv",
    primaryClass = "hep-ph",
    doi = "10.1007/JHEP05(2017)024",
    journal = "JHEP",
    volume = "05",
    pages = "024",
    year = "2017"
}

@article{Shrock:1981cq,
    author = "Shrock, Robert E.",
    title = "{Pure Leptonic Decays With Massive Neutrinos and Arbitrary Lorentz Structure}",
    reportNumber = "ITP-SB-81-31",
    doi = "10.1016/0370-2693(82)91074-7",
    journal = "Phys. Lett. B",
    volume = "112",
    pages = "382--386",
    year = "1982"
}

@article{Marquez:2022bpg,
    author = "M\'arquez, Juan Manuel and Castro, Gabriel L\'opez and Roig, Pablo",
    title = "{Michel parameters in the presence of massive Dirac and Majorana neutrinos}",
    eprint = "2208.01715",
    archivePrefix = "arXiv",
    primaryClass = "hep-ph",
    doi = "10.1007/JHEP11(2022)117",
    journal = "JHEP",
    volume = "11",
    pages = "117",
    year = "2022"
}

@article{Kayser:1982br,
    author = "Kayser, Boris",
    title = "{Majorana Neutrinos and their Electromagnetic Properties}",
    reportNumber = "SLAC-PUB-2879",
    doi = "10.1103/PhysRevD.26.1662",
    journal = "Phys. Rev. D",
    volume = "26",
    pages = "1662",
    year = "1982"
}

@article{Doi:2005pm,
    author = "Doi, Masaru and Kotani, Tsuneyuki and Nishiura, Hiroyuki",
    title = "{New parameterization in muon decay and type of neutrino}",
    eprint = "hep-ph/0502136",
    archivePrefix = "arXiv",
    doi = "10.1143/PTP.114.845",
    journal = "Prog. Theor. Phys.",
    volume = "114",
    pages = "845--871",
    year = "2005"
}

@article{Lindner:2016wff,
    author = "Lindner, Manfred and Rodejohann, Werner and Xu, Xun-Jie",
    title = "{Coherent Neutrino-Nucleus Scattering and new Neutrino Interactions}",
    eprint = "1612.04150",
    archivePrefix = "arXiv",
    primaryClass = "hep-ph",
    doi = "10.1007/JHEP03(2017)097",
    journal = "JHEP",
    volume = "03",
    pages = "097",
    year = "2017"
}

@article{MicroBooNE:2023Muon,
    author = "Abratenko, P. and others",
    collaboration = "MicroBooNE",
    title = "{Search for heavy neutral leptons in electron-positron and neutral-pion final states with the MicroBooNE detector}",
    eprint = "2310.07660",
    archivePrefix = "arXiv",
    primaryClass = "hep-ex",
    reportNumber = "FERMILAB-PUB-23-574-ND",
    month = "10",
    year = "2023"
}

@article{Blaut:2018fis,
    author = "B\l{}aut, A. and Sobk\'ow, W.",
    title = "{Neutrino elastic scattering on polarized electrons as a tool for probing the neutrino nature}",
    eprint = "1812.09828",
    archivePrefix = "arXiv",
    primaryClass = "hep-ph",
    doi = "10.1140/epjc/s10052-020-7806-0",
    journal = "Eur. Phys. J. C",
    volume = "80",
    number = "3",
    pages = "261",
    year = "2020"
}

@article{NA62:2017Electron,
    author = "Cortina Gil, Eduardo and others",
    collaboration = "NA62",
    title = "{Search for heavy neutral lepton production in $K^+$ decays}",
    eprint = "1712.00297",
    archivePrefix = "arXiv",
    primaryClass = "hep-ex",
    reportNumber = "CERN-EP-2017-311",
    doi = "10.1016/j.physletb.2018.01.031",
    journal = "Phys. Lett. B",
    volume = "778",
    pages = "137--145",
    year = "2018"
}

@article{Barouki:2022bkt,
    author = "Barouki, Ryan and Marocco, Giacomo and Sarkar, Subir",
    title = "{Blast from the past II: Constraints on heavy neutral leptons from the BEBC WA66 beam dump experiment}",
    eprint = "2208.00416",
    archivePrefix = "arXiv",
    primaryClass = "hep-ph",
    doi = "10.21468/SciPostPhys.13.5.118",
    journal = "SciPost Phys.",
    volume = "13",
    pages = "118",
    year = "2022"
}

@article{Bischer:2019ttk,
    author = "Bischer, Ingolf and Rodejohann, Werner",
    title = "{General neutrino interactions from an effective field theory perspective}",
    eprint = "1905.08699",
    archivePrefix = "arXiv",
    primaryClass = "hep-ph",
    doi = "10.1016/j.nuclphysb.2019.114746",
    journal = "Nucl. Phys. B",
    volume = "947",
    pages = "114746",
    year = "2019"
}

@article{Atre:2009rg,
    author = "Atre, Anupama and Han, Tao and Pascoli, Silvia and Zhang, Bin",
    title = "{The Search for Heavy Majorana Neutrinos}",
    eprint = "0901.3589",
    archivePrefix = "arXiv",
    primaryClass = "hep-ph",
    reportNumber = "FERMILAB-PUB-08-086-T, NSF-KITP-08-54, MADPH-06-1466, DCPT-07-198, IPPP-07-99",
    doi = "10.1088/1126-6708/2009/05/030",
    journal = "JHEP",
    volume = "05",
    pages = "030",
    year = "2009"
}

@article{Long:2014zva,
    author = "Long, Andrew J. and Lunardini, Cecilia and Sabancilar, Eray",
    title = "{Detecting non-relativistic cosmic neutrinos by capture on tritium: phenomenology and physics potential}",
    eprint = "1405.7654",
    archivePrefix = "arXiv",
    primaryClass = "hep-ph",
    doi = "10.1088/1475-7516/2014/08/038",
    journal = "JCAP",
    volume = "08",
    pages = "038",
    year = "2014"
}

@article{Lunardini:2019zob,
    author = "Lunardini, Cecilia and Perez-Gonzalez, Yuber F.",
    title = "{Dirac and Majorana neutrino signatures of primordial black holes}",
    eprint = "1910.07864",
    archivePrefix = "arXiv",
    primaryClass = "hep-ph",
    reportNumber = "FERMILAB-PUB-19-521-T, NUHEP-TH/19-14",
    doi = "10.1088/1475-7516/2020/08/014",
    journal = "JCAP",
    volume = "08",
    pages = "014",
    year = "2020"
}

@article{Luo:2020sho,
    author = "Luo, Xuheng and Rodejohann, Werner and Xu, Xun-Jie",
    title = "{Dirac neutrinos and $N_{{\rm eff}}$}",
    eprint = "2005.01629",
    archivePrefix = "arXiv",
    primaryClass = "hep-ph",
    doi = "10.1088/1475-7516/2020/06/058",
    journal = "JCAP",
    volume = "06",
    pages = "058",
    year = "2020"
}

@article{Luo:2020fdt,
    author = "Luo, Xuheng and Rodejohann, Werner and Xu, Xun-Jie",
    title = "{Dirac neutrinos and N$_{eff}$. Part II. The freeze-in case}",
    eprint = "2011.13059",
    archivePrefix = "arXiv",
    primaryClass = "hep-ph",
    doi = "10.1088/1475-7516/2021/03/082",
    journal = "JCAP",
    volume = "03",
    pages = "082",
    year = "2021"
}

@article{BahaBalantekin:2018ppj,
    author = "Baha Balantekin, A. and Kayser, Boris",
    title = "{On the Properties of Neutrinos}",
    eprint = "1805.00922",
    archivePrefix = "arXiv",
    primaryClass = "hep-ph",
    reportNumber = "FERMILAB-PUB-18-149-T",
    doi = "10.1146/annurev-nucl-101916-123044",
    journal = "Ann. Rev. Nucl. Part. Sci.",
    volume = "68",
    pages = "313--338",
    year = "2018"
}

@article{Balantekin:2018ukw,
    author = "Balantekin, A. Baha and de Gouv\^ea, Andr\'e and Kayser, Boris",
    title = "{Addressing the Majorana vs. Dirac Question with Neutrino Decays}",
    eprint = "1808.10518",
    archivePrefix = "arXiv",
    primaryClass = "hep-ph",
    reportNumber = "FERMILAB-PUB-18-418-T, NUHEP-TH/18-09",
    doi = "10.1016/j.physletb.2018.11.068",
    journal = "Phys. Lett. B",
    volume = "789",
    pages = "488--495",
    year = "2019"
}

@article{Funcke:2019grs,
    author = "Funcke, Lena and Raffelt, Georg and Vitagliano, Edoardo",
    title = "{Distinguishing Dirac and Majorana neutrinos by their decays via Nambu-Goldstone bosons in the gravitational-anomaly model of neutrino masses}",
    eprint = "1905.01264",
    archivePrefix = "arXiv",
    primaryClass = "hep-ph",
    reportNumber = "MPP-2019-67",
    doi = "10.1103/PhysRevD.101.015025",
    journal = "Phys. Rev. D",
    volume = "101",
    number = "1",
    pages = "015025",
    year = "2020"
}

@article{SajjadAthar:2021prg,
    author = "Sajjad Athar, Mohammad and others",
    title = "{Status and perspectives of neutrino physics}",
    eprint = "2111.07586",
    archivePrefix = "arXiv",
    primaryClass = "hep-ph",
    reportNumber = "FERMILAB-PUB-21-621-ND",
    doi = "10.1016/j.ppnp.2022.103947",
    journal = "Prog. Part. Nucl. Phys.",
    volume = "124",
    pages = "103947",
    year = "2022"
}
\end{document}